\documentclass[a4paper]{jpconf}

\pdfoutput=1

\usepackage{graphicx}

\def\cpkkd{\rm{kg^{-1}keV^{-1}day^{-1}}}

\def\s2tw{\rm{ sin ^2 \theta _W }}

\begin{document}
\title{Dark Matter Search
with sub-keV Germanium Detectors
at the China Jinping Underground Laboratory
}

\author{Qian Yue $^a$ and Henry T. Wong $^b$\\
(on behalf of the CDEX-TEXONO Collaboration$^{\dagger}$)
}

\address{
$^a$ Department of Engineering Physics, Tsinghua University,
Beijing 100084, China.\\
$^b$ Institute of Physics, Academia Sinica,
Taipei 11529, Taiwan.
}

\ead{$^a$ yueq@mail.tsinghua.edu.cn ; $^b$ htwong@phys.sinica.edu.tw}

\begin{abstract}

Germanium detectors with sub-keV 
sensitivities open a window
to search for low-mass WIMP dark matter.
The CDEX-TEXONO Collaboration
is conducting the first research 
program at the new 
China Jinping Underground Laboratory
with this approach.
The status and plans
of the laboratory and the experiment
are discussed.

\end{abstract}



The theme of
the CDEX-TEXONO research program is
on the studies of
low energy neutrino and dark matter physics.
The current objectives are to 
open the ``sub-keV'' detector window
with germanium detectors~\cite{jpcs06}.
The generic ``benchmark''
goals in terms of detector performance are:
(1) modular target mass of order of 1~kg;
(2) detector sensitivities reaching the range of 100~eV;
(3) background at the range of $1 ~ \cpkkd$ (cpkkd).
The neutrino physics program~\cite{texononuphys} 
is pursued at the established
Kuo-Sheng Reactor Neutrino Laboratory (KSNL),
while dark matter searches  will be conducted
at the new China Jin-Ping Underground
Laboratory (CJPL)~\cite{cjplnews}
officially inaugurated in December 2010.
The three main scientific subjects are
neutrino magnetic moments,
neutrino-nucleus coherent scattering,
and dark matter searches.
We highlight the status and plans
of the dark matter program 
in this article.

There are compelling evidence that about one-quarter 
of the energy density in the universe is
composed of Cold Dark Matter~\cite{cdmpdg10} due to
a not-yet-identified particle, generically categorized
as Weakly Interacting Massive Particle
(WIMP, denoted by $\chi$).
A direct experimental detection of WIMP
is one of the biggest challenges in the
frontiers of particle physics and cosmology.
The WIMPs interact with matter pre-dominantly
via elastic coherent scattering
like the neutrinos:
$\rm{
\chi  +  N  \rightarrow 
\chi  +  N ~ .
}$
There may be both
spin-independent ($\sigma ^{SI} _{\chi N}$) 
and spin-dependent ($\sigma ^{SD} _{\chi N}$) 
interactions between WIMP and matter.

\begin{figure}
\begin{minipage}{8cm}
\includegraphics[width=8cm]{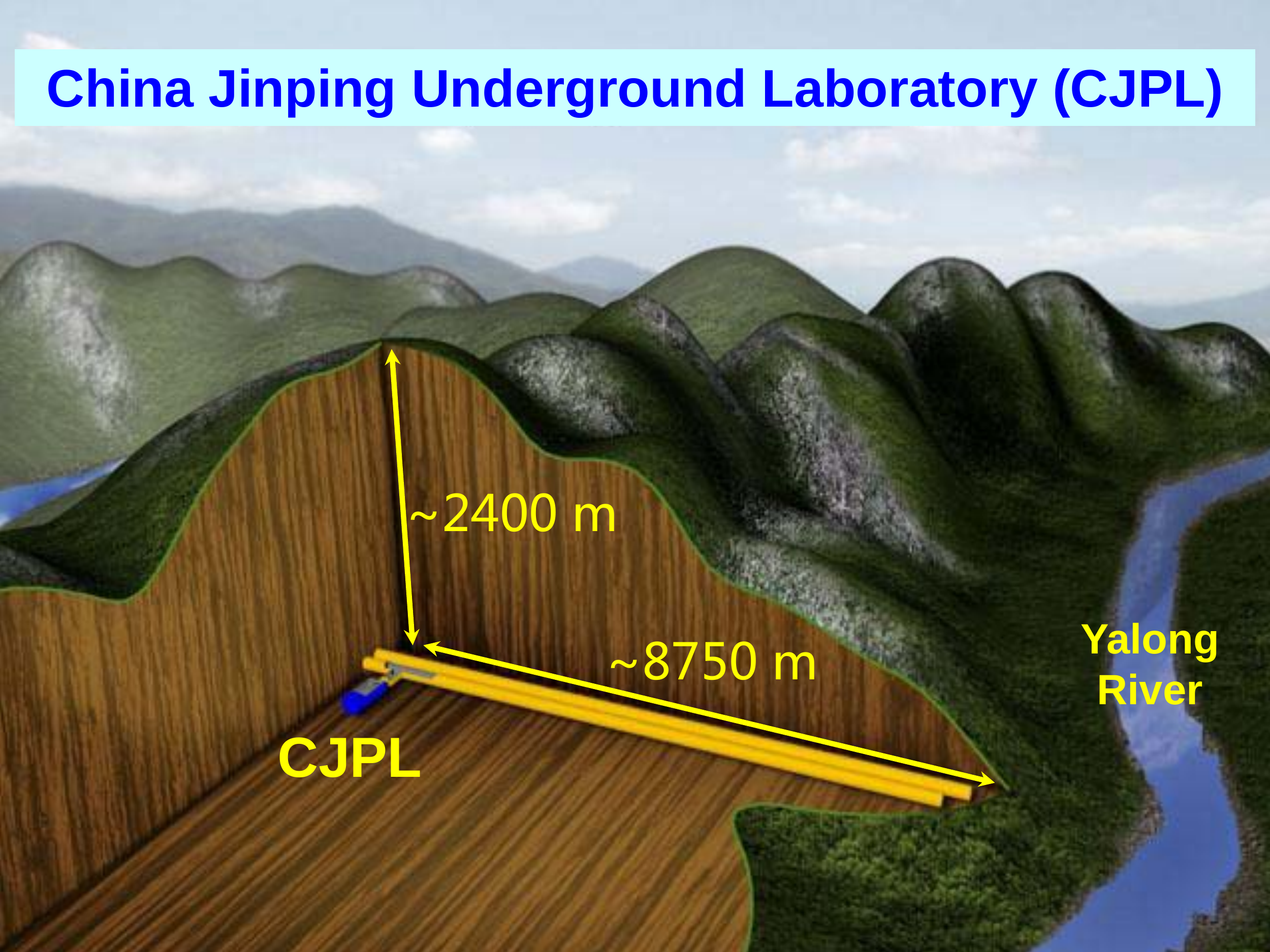}
\end{minipage}
\hfill
\begin{minipage}{8cm}
\includegraphics[width=8cm]{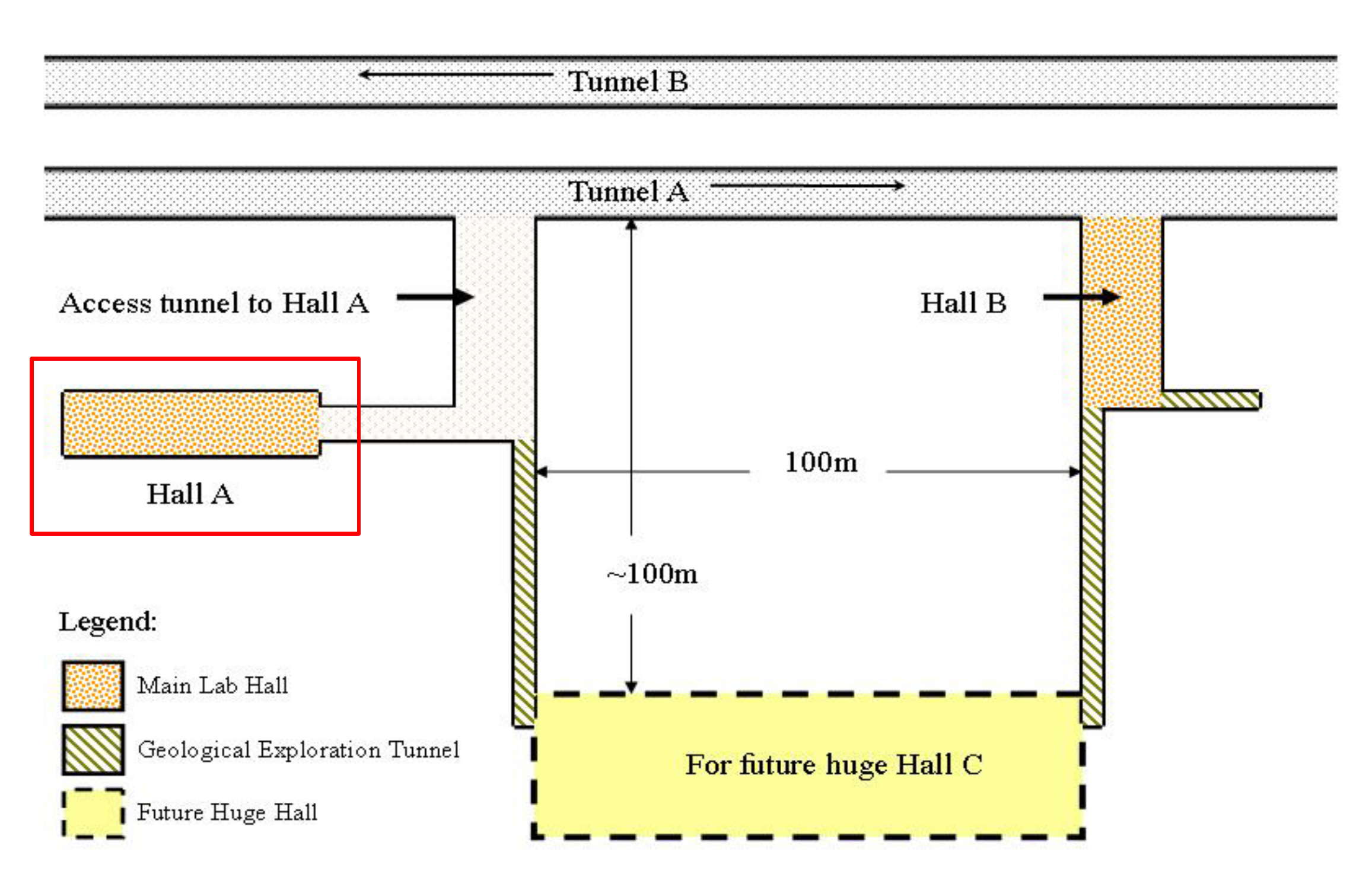}
\includegraphics[width=8cm]{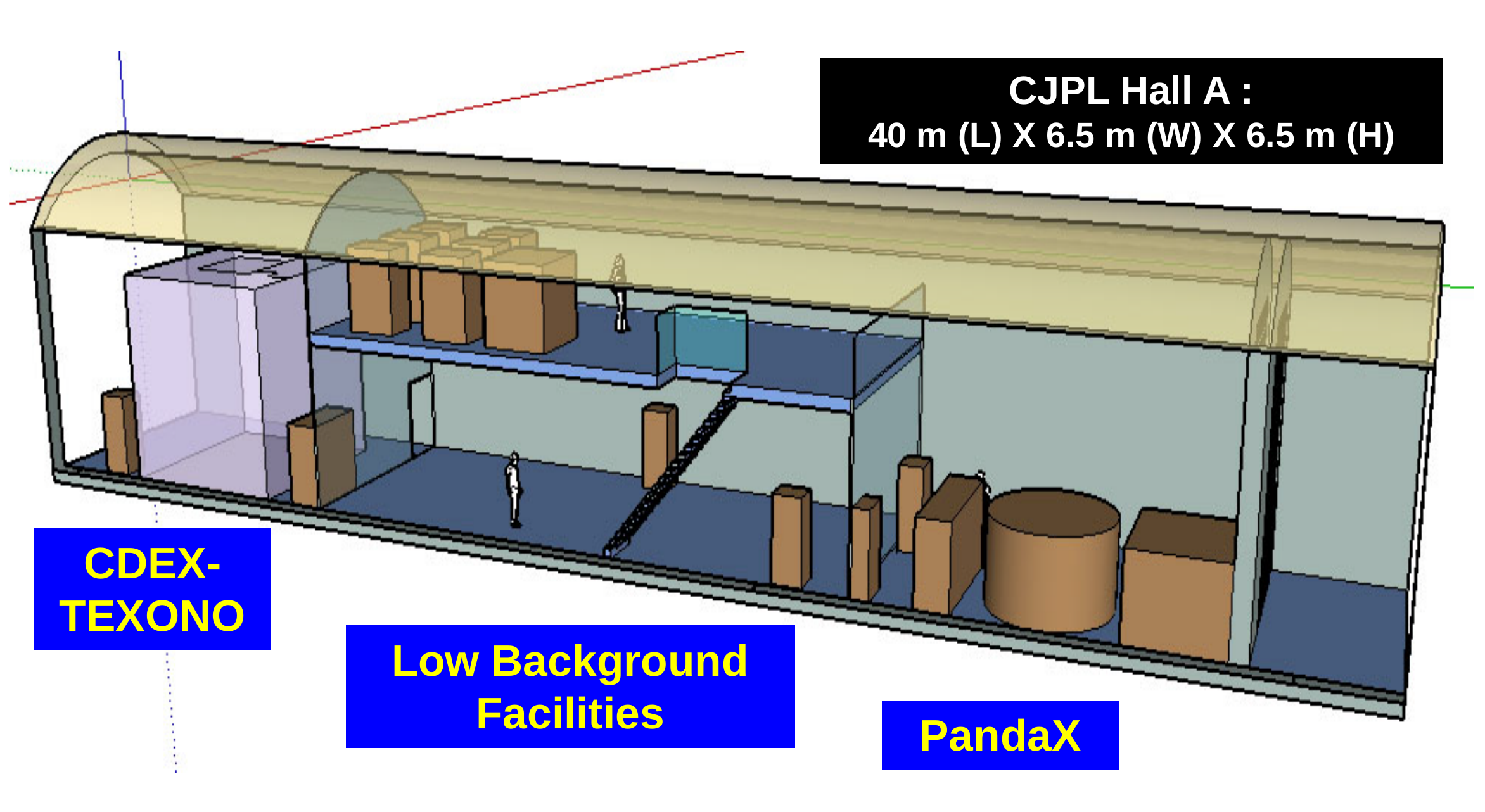}
\end{minipage}
\caption{
Schematics diagrams displaying the essential features of CJPL:
Left$-$ Geographical setting showing tunnel length and
overburden; 
Right Top$-$ Floor Plan  
of the present Hall-A and expected future expansions;
Right Bottom$-$ Layout of Hall-A, showing
both CDEX-TEXONO and PandaX experiments.
}
\label{cjpl}
\end{figure}

The facility CJPL~\cite{cjplnews}
is the deepest operating underground 
laboratory in the world,
having $\sim$2400 meter of rock overburden 
and tunnel drive-in access,
as shown schematically in Figure~\ref{cjpl},
It is located at southwest Sichuan, China,
reachable from the provincial international airport
at Chengdu via a 50~min flight to Xichang followed by
a 90~min drive on a private two-lane motorway.
The laboratory is owned by 
the Ertan Hydropower Development Company,
and managed by Tsinghua University, China.
Excavation and construction 
of the first experimental hall (``Hall A'')
of dimension 6.5 m(width)X6.5 m(height)X40 m(length)
with 50~cm of concrete lining
was completed in summer 2010.
By Fall 2011, 
the ventilation system,
high-speed internet connections,
as well as 
the necessary surface infrastructures
(office and dormitory spaces,
liquid nitrogen storage system) 
have been installed.
There are intense efforts at CJPL
to characterize the background. 
Measurements are being performed on the
ambient radioactivity as well
as fast and thermal neutron fluxes.
Residual cosmic-ray events have been observed,
at a rate (several events per month per square-meter)
consistent with the expectation for a location with
2400~m rock overburden.
The first generation experimental program at CJPL
will include two projects: the CDEX-TEXONO experiment
described here, and the 
dark matter project PandaX
with liquid xenon detector.
Future expansions of the laboratory are foreseen.
New ideas are being discussed and explored.

\begin{figure}
\begin{minipage}{8cm}
\includegraphics[width=8cm]{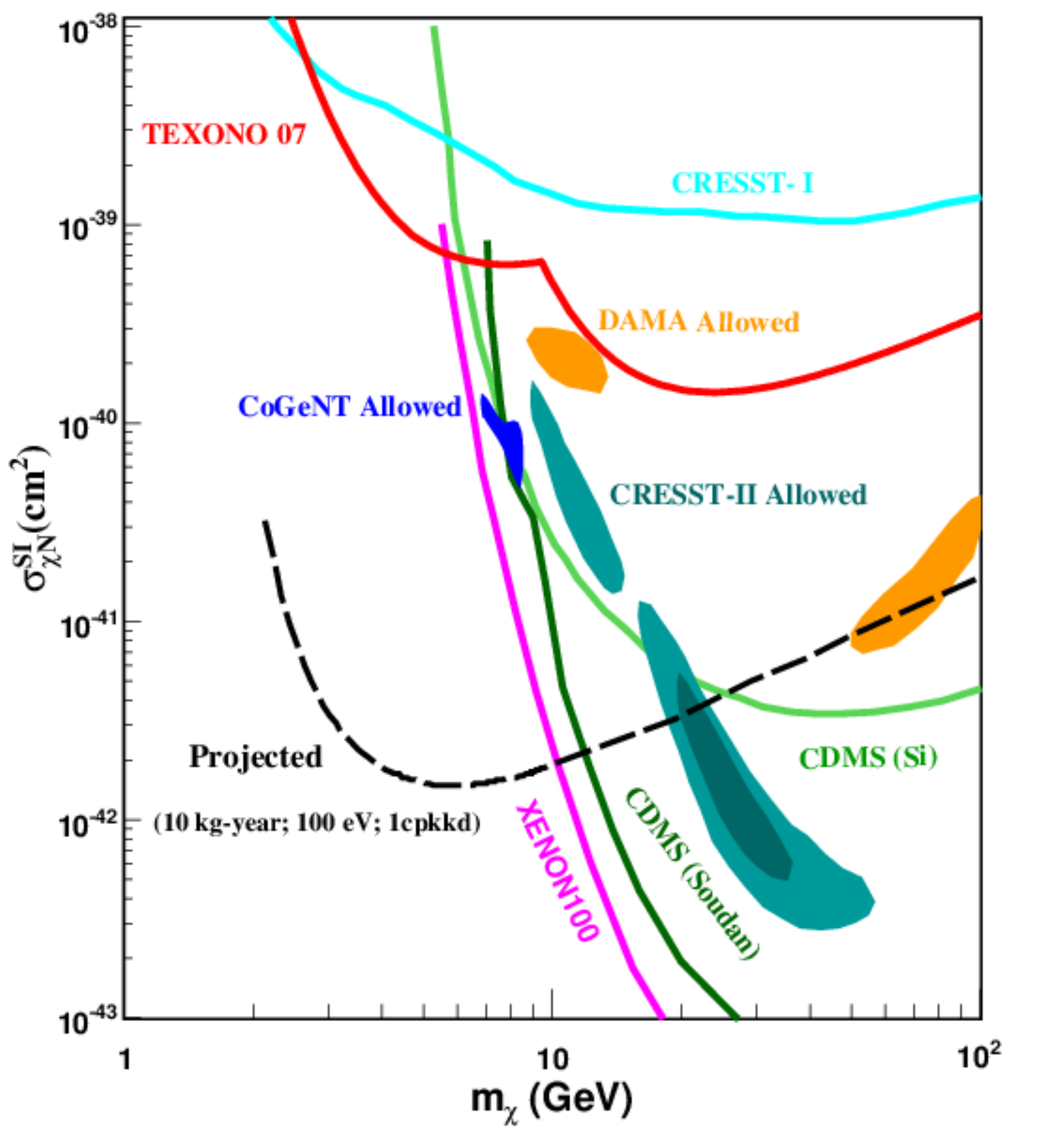}
\end{minipage}
\hfill
\begin{minipage}{8cm}
\includegraphics[width=8cm]{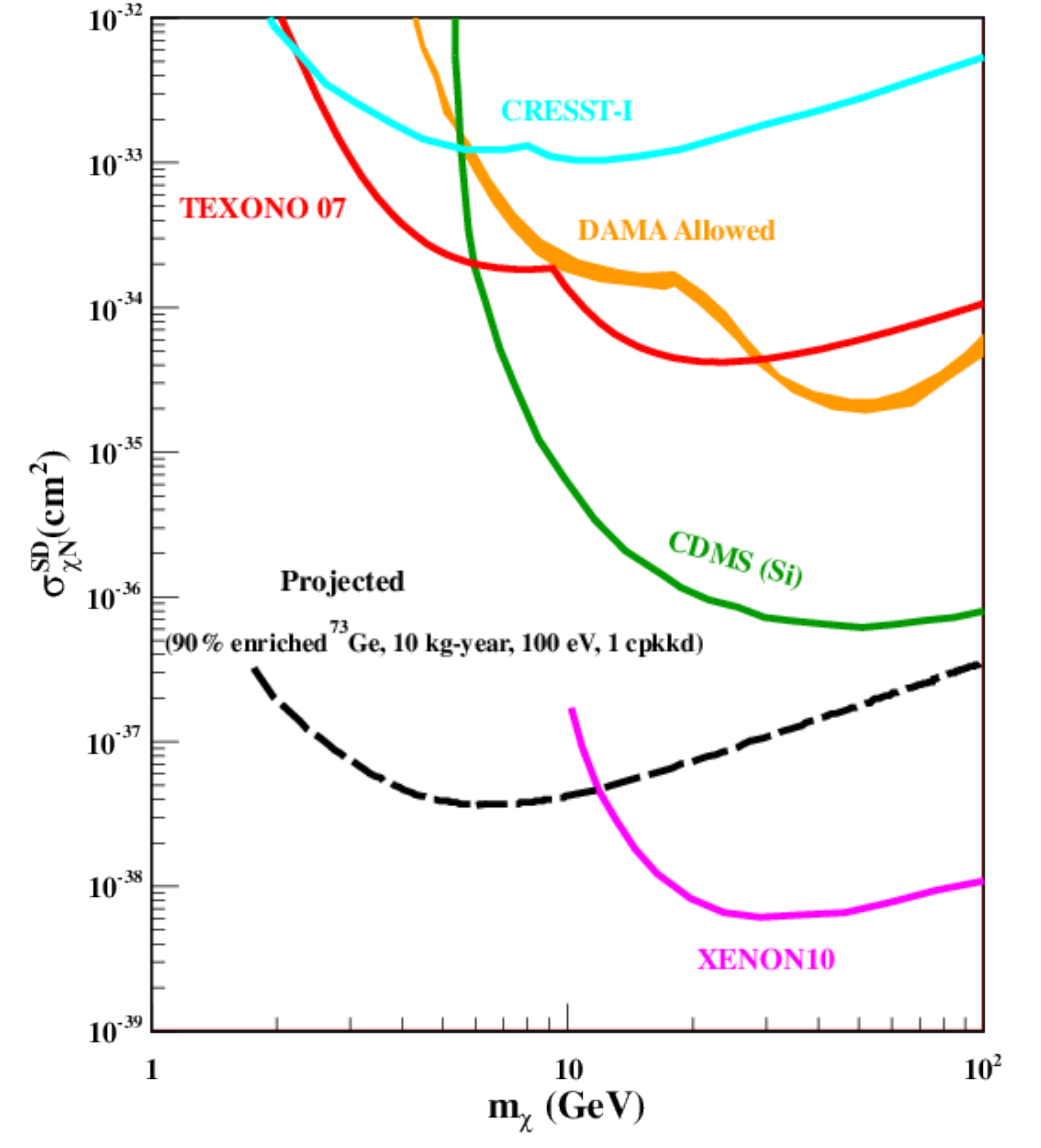}
\end{minipage}
\caption{
Exclusion plots of 
(a) Left: spin-independent $\chi$N 
and
(b) Right: spin-dependent $\chi$N 
cross-sections versus
WIMP-mass, displaying the
KSNL-ULEGe limits~\cite{texonocdm07} and
those defining the
current boundaries~\cite{cdmpdg10,taup11}.
The DAMA, CoGeNT and CRESST-II 
allowed regions~\cite{taup11,cogent}
are superimposed.
Projected reach of
experiments at benchmark sensitivities
are indicated as dotted lines.
}
\label{explot}
\end{figure}

An experiment with 100~eV threshold would
open a window for Cold Dark Matter
WIMP searches~\cite{cdmpdg10}
in the unexplored mass range
down to several GeV~\cite{jpcs06}.
Based on data taken at KSNL with the 20-g prototype 
Ultra-Low-Energy Germanium detector (ULEGe),
limits were derived in this low WIMP mass region 
improving over those from the previous experiments at 
$ 3 < m_\chi < 6 ~{\rm GeV}$~\cite{texonocdm07}.
The
$\rm{ \sigma ^{SI} _{\chi N} }$ versus $m_{\chi}$ 
and 
$\rm{ \sigma ^{SD} _{\chi N} }$ versus $m_{\chi}$ 
exclusion plots are 
depicted in Figures~\ref{explot}a\&b, respectively.
Also displayed are the various results
defining the exclusion boundaries,
together with allowed regions implied
by the DAMA/LIBRA, CoGeNT and CRESST-II data~\cite{taup11,cogent}.
In particular, interpretations of the recent 
CoGeNT low-energy spectra as positive signatures
of low-mass WIMPs~\cite{cogent} have stimulated
intense theoretical interests and speculations
on this parameter space.

Point-Contact Germanium detectors (PCGe)~\cite{pcge}
offer sub-keV sensitivities with 
detector of kg-size modular mass,
an improvement over the conventional ULEGe design.
WIMPs with mass down to a few GeV can be probed.
Intensive R\&D efforts~\cite{pcgerandd}
are pursued to optimize 
the application of PCGe in dark matter searches,
including programs on: 
(1) pulse shape analysis of near noise-edge events
to extend the physics range, 
(2) pulse shape analysis of surface versus bulk events 
to characterize an important background channel,
(3) sub-keV background understanding and suppression, and
(4) fabrication of advanced electronics for Ge detectors. 

A polyethylene (PE) shielding structure with thickness 1~m 
and interior dimension 
8~m(length)$\times$4.5~m(width)$\times$4~m(height) 
has been constructed for the CDEX-TEXONO program at CJPL.
A 20-g ULEGe array and  
a 1-kg PCGe have been installed  
within OFHC copper shielding inside this PE-housing.
Data taking has commenced in February 2011.
Design and construction of the next-generation
PCGe array with total mass at the 10-kg range 
is proceeding.
This new detector will be shielded and 
enclosed in a liquid argon chamber 
which serves as both
cryogenic medium and active shielding and 
anti-Compton detector where the
scintillation light will be read out
by photomultipliers.
Commissioning is planned in 2013.
Potential reaches 
are depicted by dotted lines in 
Figures~\ref{explot}a\&b.
The projected sensitivities assume
Ge detectors at 100~eV threshold 
(equivalent to about 500~eV nuclear recoils),
10 kg-year of exposure and that 
the achieved background level of 
the order of 1~cpkkd at the few keV range 
can be extrapolated down to threshold.

$^\dagger$ The CDEX-TEXONO Collaboration
consists of groups from
China (Tsinghua University, China Institute of Atomic Energy,
Nankai University, Sichuan University, Ertan Hydropower Development Company),
Taiwan (Academia Sinica, Institute of Nuclear Energy, 
Kuo-Sheng Nuclear Power Station, National Tsing-Hua University),
India (Banaras Hindu University) and 
Turkey (Middle East Technical University, Karadeniz Technical University).
The authors are grateful to the
generous and timely support provided 
by the participating institutes and their respective
funding agencies.

\end{document}